\documentclass[10pt]{iopart}
\usepackage{dcolumn,graphicx,color,booktabs,microtype,afterpage}
\usepackage{float}
\usepackage{amssymb}
\usepackage{nicefrac}
\usepackage{url}  
\usepackage[charter,greekuppercase=italicized]{mathdesign}
\usepackage{sidecap}
\usepackage[mathlines]{lineno}
\usepackage{microtype}

\usepackage{color}
\usepackage[colorlinks,plainpages=false,linkcolor=blue,urlcolor=blue,citecolor=blue,pdfpagemode=UseNone,pdfstartview=FitBH]{hyperref}
\graphicspath{{./}{figures/}}

\usepackage[dvipsnames]{xcolor}

\newcommand{\tcr}[1]{\textcolor{black}{#1}}
\newcommand{\tcb}[1]{\textcolor{black}{#1}}

\newcommand{\tcg}[1]{\textcolor{black}{#1}}
\newcommand{\tco}[1]{\textcolor{black}{#1}}

\begin{document}
	\makeatletter\renewcommand{\ps@plain}{%
		\def\@evenhead{\hfill\itshape\rightmark}%
		\def\@oddhead{\itshape\leftmark\hfill}%
		\renewcommand{\@evenfoot}{\hfill\small{--~\thepage~--}\hfill}%
		\renewcommand{\@oddfoot}{\hfill\small{--~\thepage~--}\hfill}%
	}\makeatother\pagestyle{plain}
\title[NMR evidence of $s$-wave superconductivity in \texorpdfstring{W$_3$Al$_2$C}{W3Al2C}]{$s$-wave superconductivity in the noncentrosymmetric \texorpdfstring{W$_3$Al$_2$C}{W3Al2C} superconductor: 
An NMR study}
\author{D Tay$^{1}$, T Shang$^{2*}$, Y P Qi$^{3}$, T P Ying$^{4,5}$, 
        H Hosono$^{4}$, H-R Ott$^{1,6}$, and T~Shiroka$^{1,6*}$}
\address{$^1$Laboratorium f\"ur Festk\"orperphysik, ETH Z\"urich, CH-8093 Zurich, Switzerland}
\address{$^2$Key Laboratory of Polar Materials and Devices (MOE), School of Physics 
             and Electronic Science, East China Normal University, Shanghai 200241, China}
\address{$^3$School of Physical Science and Technology, ShanghaiTech University, Shanghai 201210, China} 
\address{$^4$Materials Research Center for Element Strategy, Tokyo Institute of Technology, Yokohama 226-8503, Japan}
\address{$^5$Beijing National Laboratory for Condensed Matter Physics, Institute of Physics, Chinese Academy of Sciences, Beijing 100190, China}
\address{$^6$Paul Scherrer Institut, CH-5232 Villigen PSI, Switzerland}
\eads{\mailto{tshiroka@phys.ethz.ch}, \mailto{tshang@phy.ecnu.edu.cn}}
\date{\today}
	
\begin{abstract}
We report on a microscopic study of the noncentrosymmetric superconductor 
W$_3$Al$_2$C (with $T_c = 7.6$\,K), mostly by means of $^{27}$Al- and
$^{13}$C nuclear magnetic resonance (NMR). Since in this
material the density of states at the Fermi level is dominated by the
tungsten's 5$d$ orbitals, we expect a sizeable spin-orbit coupling
(SOC) effect. The normal-state electronic properties of W$_3$Al$_2$C
resemble those of a standard metal, but with a Korringa product
$1/(T_{1}T)$ significantly smaller than that of metallic Al, 
reflecting the marginal role played by $s$-electrons. 
In the superconducting state, we observe a reduction of the Knight shift
and an exponential decrease of the NMR relaxation rate $1/T_1$, typical
of $s$-wave superconductivity. 
This is further \tco{supported} 
by the observation of a \tco{small but distinct} coherence peak just
below $T_c$ in the $^{13}$C NMR relaxation-rate, in agreement with the
fully-gapped superconducting state inferred from the electronic
specific-heat data well below $T_c$. 
The above features are compared to those of members of the same family, 
in particular, Mo$_3$Al$_2$C, \tco{often claimed to exhibit} 
unconventional superconductivity.  
We discuss why, despite the enhanced SOC, W$_3$Al$_2$C does not show
spin-triplet features in its superconducting state and consider the
broader consequences \tco{of our results} for noncentrosymmetric superconductors in general.
\end{abstract}

\vspace{2pc}

\submitto{\JPCM}
%
\maketitle
\ioptwocol
\maketitle\enlargethispage{3pt}

\vspace{-5pt}
\section{\label{sec:Introduction}Introduction}
Non\-cen\-tro\-sym\-met\-ric superconductors (NCSCs) belong to a class 
of materials that miss a key symmetry, such as parity~\cite{Bauer2012}. 
In NCSCs, the lack of inversion symmetry of the crystal lattice often 
induces an antisymmetric spin-orbit coupling (ASOC), which lifts the 
degeneracy of the conduction-electron bands and splits the Fermi surface. 
Consequently, both intra- and inter-band Cooper pairs can be formed and 
an admixture of spin-singlet and spin-triplet pairings is possible. 
The extent of the sin\-glet\--trip\-let admixture is notably determined
not just by the strength of the ASOC, but also by other microscopic
parameters~\cite{Bauer2012,Smidman2017}. 
\textcolor{black}{In superconductors with spin-triplet pairing, time-reversal 
symmetry (TRS) breaking is not a strict requirement~\cite{Anderson1984}.
\tcg{Hence, they are particularly interesting for studying TRS, 
should it occur~\cite{Bauer2012}.}}
A notable example is that of UPt$_3$, where
claims of triplet superconductivity (SC) have been confirmed by various
experimental methods~\cite{Joynt2002}, including measurements of zero-field
muon-spin relaxation~\cite{Luke1993} and optical Kerr effect~\cite{Schemm2014}.
%

A particular NCSC family, which exhibits unconventional superconducting
properties, is that of $M_3X_2Y$ compounds. For the definition of
``unconventional’' superconductor (basically one that does not conform
to $s$-wave pairing), we follow the definition put forward in
ref.~\cite{Norman2011}. Here, $M$ = Mo, Pd, W, Pt; $X$ = Li, Al; and
$Y$ = B, C, N. Five members of this family have already been studied and,
in ascending order of atomic number $Z$ of the metal atom $M$, they are:
Rh$_{2}$Mo$_{3}$N, Mo$_3$Al$_2$C, Li$_2$Pt$_3$B, W$_3$Al$_2$C, and
Li$_2$Pd$_3$B. From general principles, it is expected that as $Z$
increases, the ASOC strength increases as well. Consequently, a
superconductor of an unconventional type is more likely to appear in
metals with a higher atomic number.
Indeed, from the existing literature, it is known that, while Li$_2$Pd$_3$B
is a conventional superconductor \cite{nishiyama2005superconductivity}, 
its high-$Z$ Pt counterpart, Li$_2$Pt$_3$B, exhibits unconventional 
superconductivity~\cite{shimamura2007nmr}, here identified by the
presence of gap nodes~\cite{Yuan2006}. \textcolor{black}{
The superconducting properties of the \tco{isostructural} 
Mo$_3$Al$_2$C vs W$_3$Al$_2$C \tco{compounds} 
are, however, still under study, with various groups}  
\tcg{reporting 
clearly \tco{contradictory} 
types of superconductivity, for either of them, 
both conventional~\cite{Ying2019,Gupta2021,karki2010structure,Koyama2013,koyama2011partial,bonalde2011evidence,zhigadlo2018crystal} and unconventional~\cite{bauer2010unconventional,Kuo2012}.}
\tcb{Thus, in Mo$_3$Al$_2$C, a power-law behavior of the $^{27}$Al NMR 
spin-lattice relaxation rates possibly suggests superconducting gap nodes~\cite{bauer2010unconventional}, 
while the exponential temperature dependence of 
the magnetic penetration depth and the absence of time-reversal symmetry 
breaking are more consistent with conventional nodeless SC~\cite{bonalde2011evidence,bauer2014absence}.} 
\tcr{More experimental data is required to fully \tco{establish} 
the behavior of these systems.}

In this paper, we explore the electronic properties of W$_3$Al$_2$C,
in both the normal- and the superconducting states. 
Our \tcg{results indicate} that this system is only weakly correlated. 
Most importantly, based on our NMR \tcg{experiments, we provide} 
evidence of BCS-type $s$-wave pairing in the noncentrosymmetric 
W$_3$Al$_2$C superconductor. \textcolor{black}{It is surprising that 
unconventional superconductivity is 
\tco{not observed,} even though W
has a higher atomic number than Mo (and, hence, a larger ASOC). We argue why, despite the presence of the heavier
element, W$_3$Al$_2$C exhibits only conventional SC behaviour.}

\begin{figure*}[htbp]
\centering 
   \includegraphics[width=0.49\textwidth]{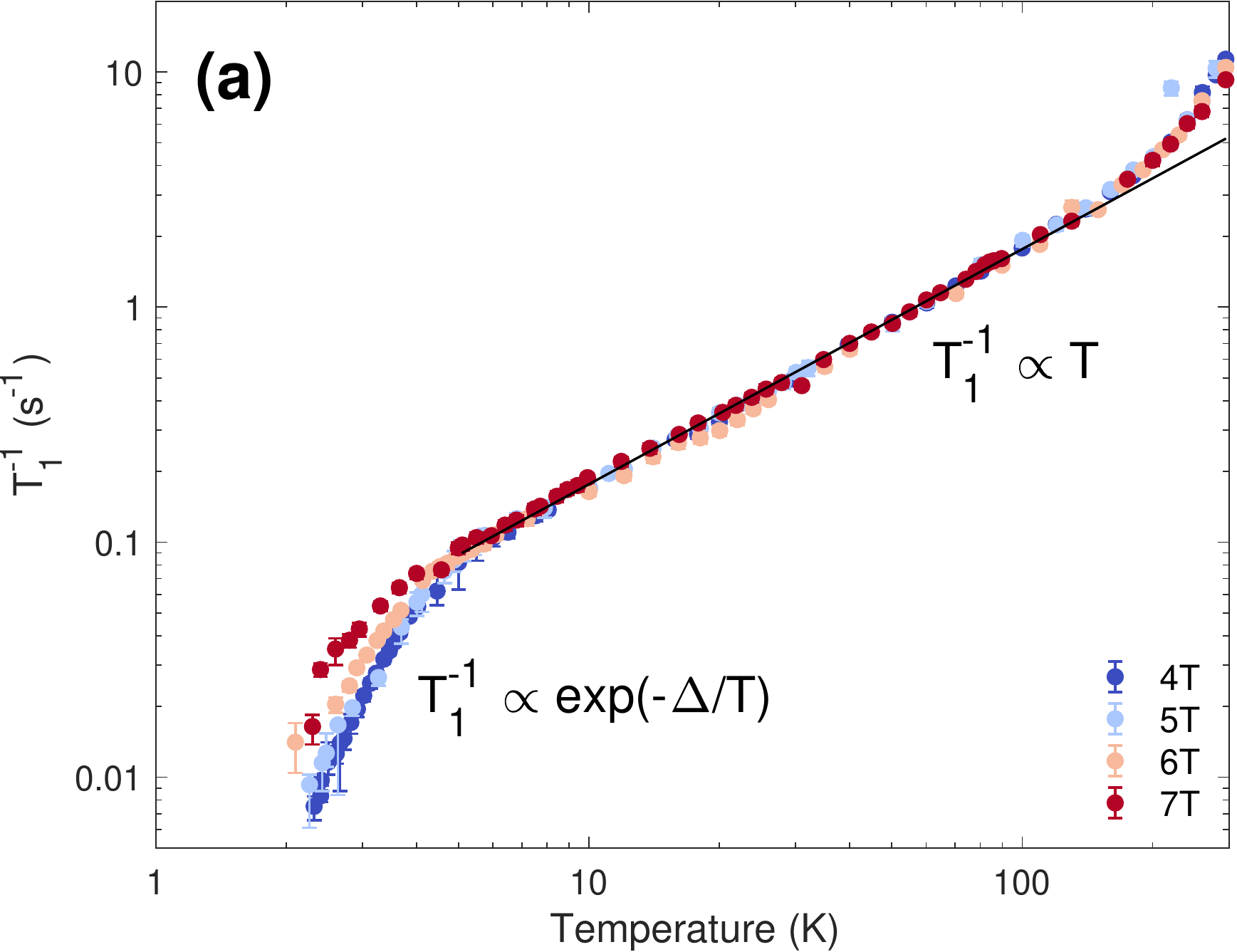}
   \hspace*{5mm}
   \includegraphics[width=0.495\textwidth]{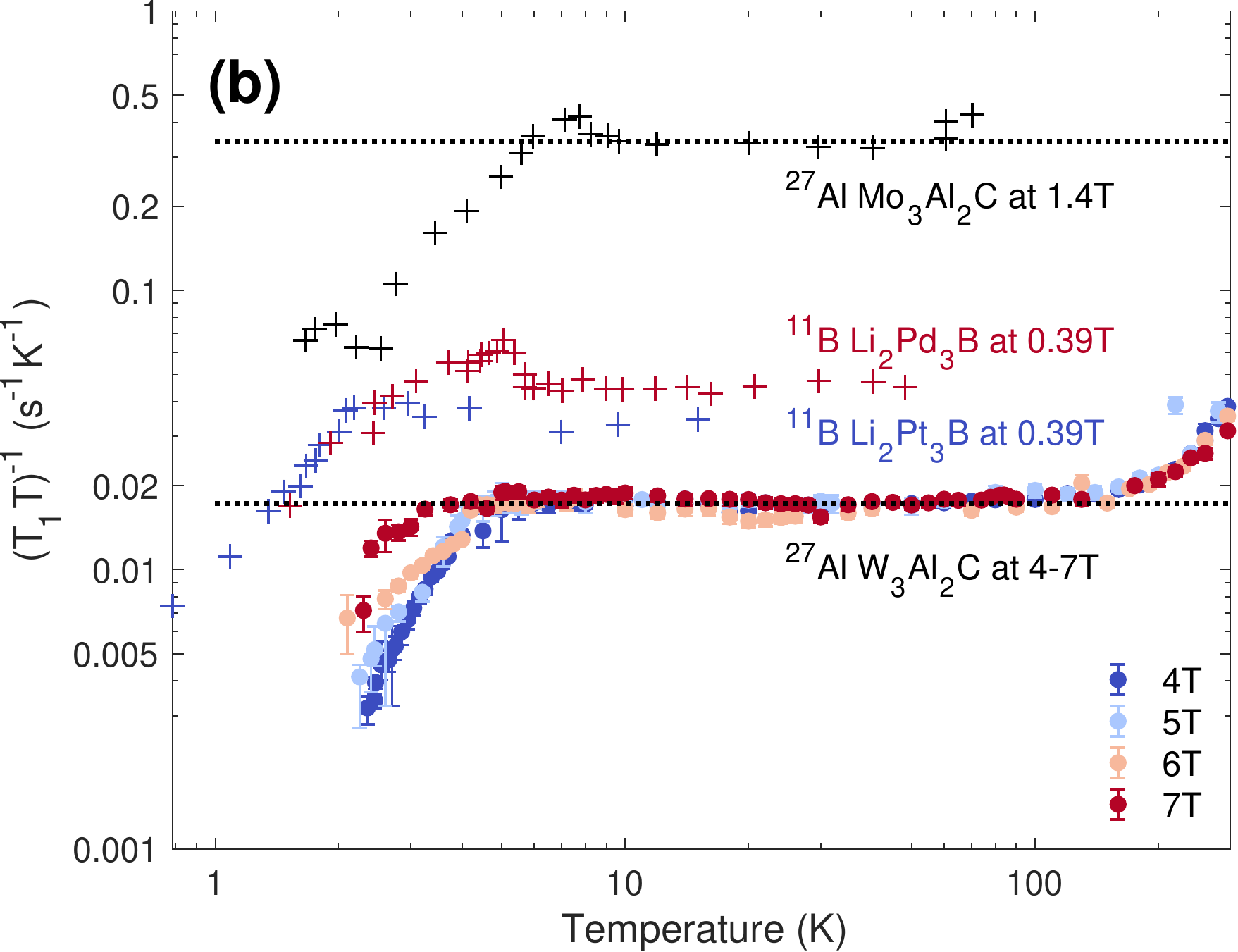}
   \caption{\label{fig:T1T_all}(a) $1/T_1$ vs temperature measured in
   different fields. Above $T_c$, the field independent spin-lattice
   relaxation rate is linear in temperature and changes slope at
   $T^\star \sim 103$\,K. (b) $1/T_1T$ vs temperature. For $T > T_c$
   (here 5.3\,K), the product is almost constant (0.017\,s$^{-1}$K$^{-1}$).
   \textcolor{black}{For comparison, we also show the data for Mo$_3$Al$_2$C~\cite{Koyama2013},  Li$_2$Pd$_3$B~\cite{Nishiyama2007} and
   Li$_2$Pt$_3$B~\cite{Nishiyama2007} --- see text for the discussion.}}
\end{figure*}

\section{\label{sec:details}Experimental details}
Polycrystalline samples of W$_3$Al$_2$C were prepared via high-temperature,
high-pressure solid-state reaction. To achieve a homogeneous mixture,
high-purity W, Al, and C powders in a quasi-stoichiometric ratio 3:\-1.8:\-0.8
were ball-milled for two days in a glove box under Ar atmosphere. The mixture
was then pressed into a pellet and placed in an $h$-BN capsule, where
the sample was heated up to 2173\,K under 5\,GPa for 24 hours.
Room-temperature x-ray powder diffraction (XRD) measurements were
performed in a Bruker D8 diffractometer with Cu K$\alpha$ radiation.
Rietveld refinements via the \textsc{FullProf} suite confirm the cubic
structure of W$_3$Al$_2$C, with a space group $P4_{1}32$ (No.\ 213)~\cite{Carvajal1993}.
Magnetic susceptibility and heat-capacity measurements were performed
on, respectively, a vibrating sample magnetometer (VSM) and a physical
property measurement system (PPMS), both by Quantum Design. Further details
on sample preparation and characterization have been reported in~\cite{Ying2019}.

The $^{27}$Al NMR measurements, including lineshapes and spin-lattice 
relaxation times, were performed on W$_3$Al$_2$C in powder form in
different applied magnetic fields (4--7\,T). To cover the 1.8 to 300\,K
temperature range we used a continuous-flow CF-1200 cryostat by Oxford
Instruments, with temperatures below 4.2\,K being achieved under pumped
$^{4}$He conditions. Preliminary resonance detuning experiments confirmed
a $T_{c}$ of 7.6\,K at 0\,T (5.3\,K at 5\,T). The $^{27}$Al NMR signal
was detected by means of a standard spin-echo sequence, consisting of
$\pi/2$ and $\pi$ pulses of 3 and 6\,$\mu$s, with recycling delays ranging
from 1 to 60\,s, in the 1.8--300\,K temperature range. Despite an echo
delay of 100\,$\mu$s, 2 to 32 scans were sufficient to acquire a
good-quality signal. The lineshapes were obtained via fast Fourier
transform (FFT) of the echo signal. Spin-lattice relaxation times $T_1$
were measured via the inversion recovery method, using a $\pi$--$\pi/2$--$\pi$ 
pulse sequence. Subsequently, the $T_1$ values for the central transition
of the spin-\nicefrac{5}{2} $^{27}$Al nucleus were obtained by using the relevant formula for the relaxation of quadrupole nuclei \cite{McDowell1995}.

\begin{figure*}[htbp]
\centering 
   \includegraphics[width=0.49\textwidth]{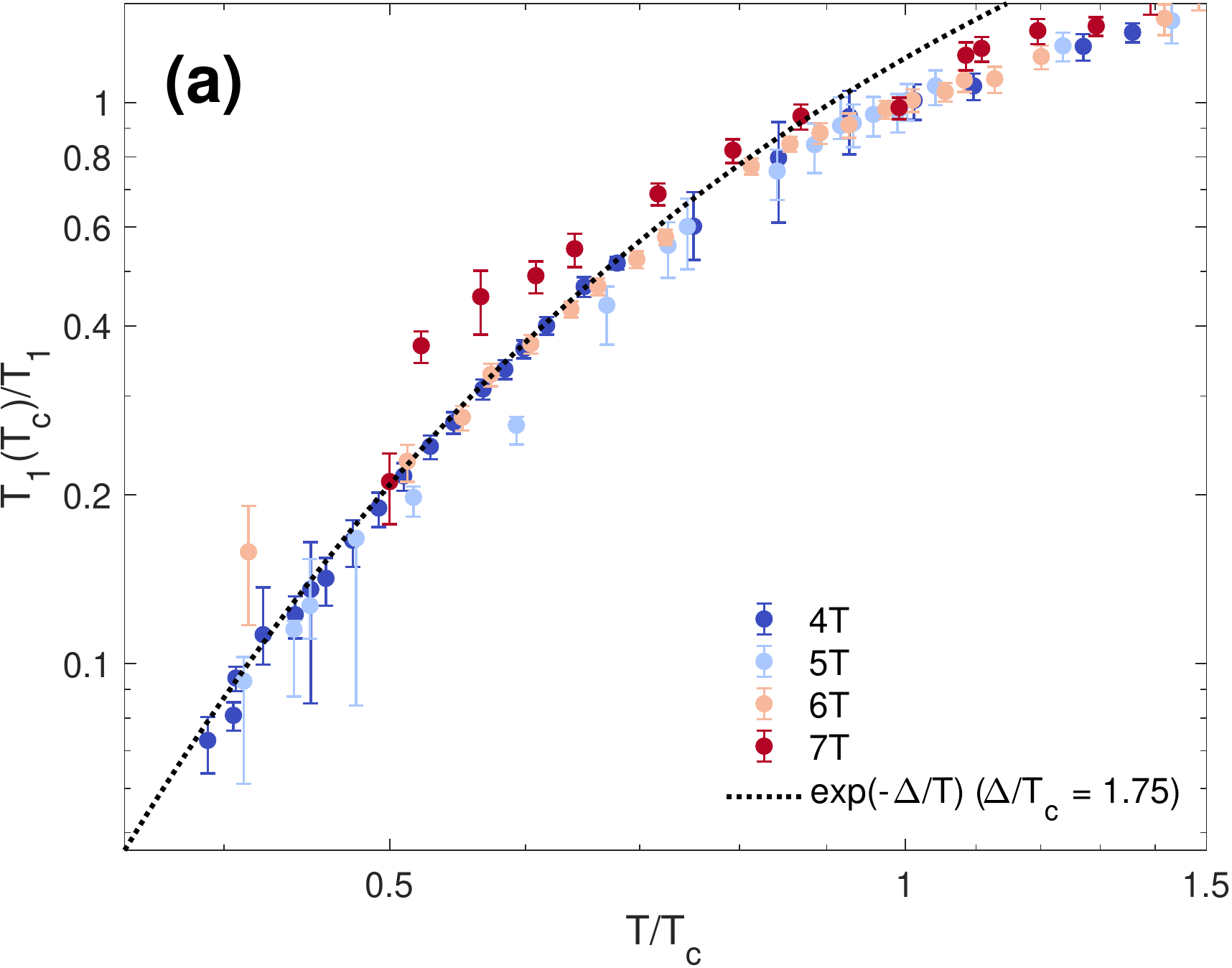}
   \hspace*{5mm}
   \includegraphics[width=0.505\textwidth]{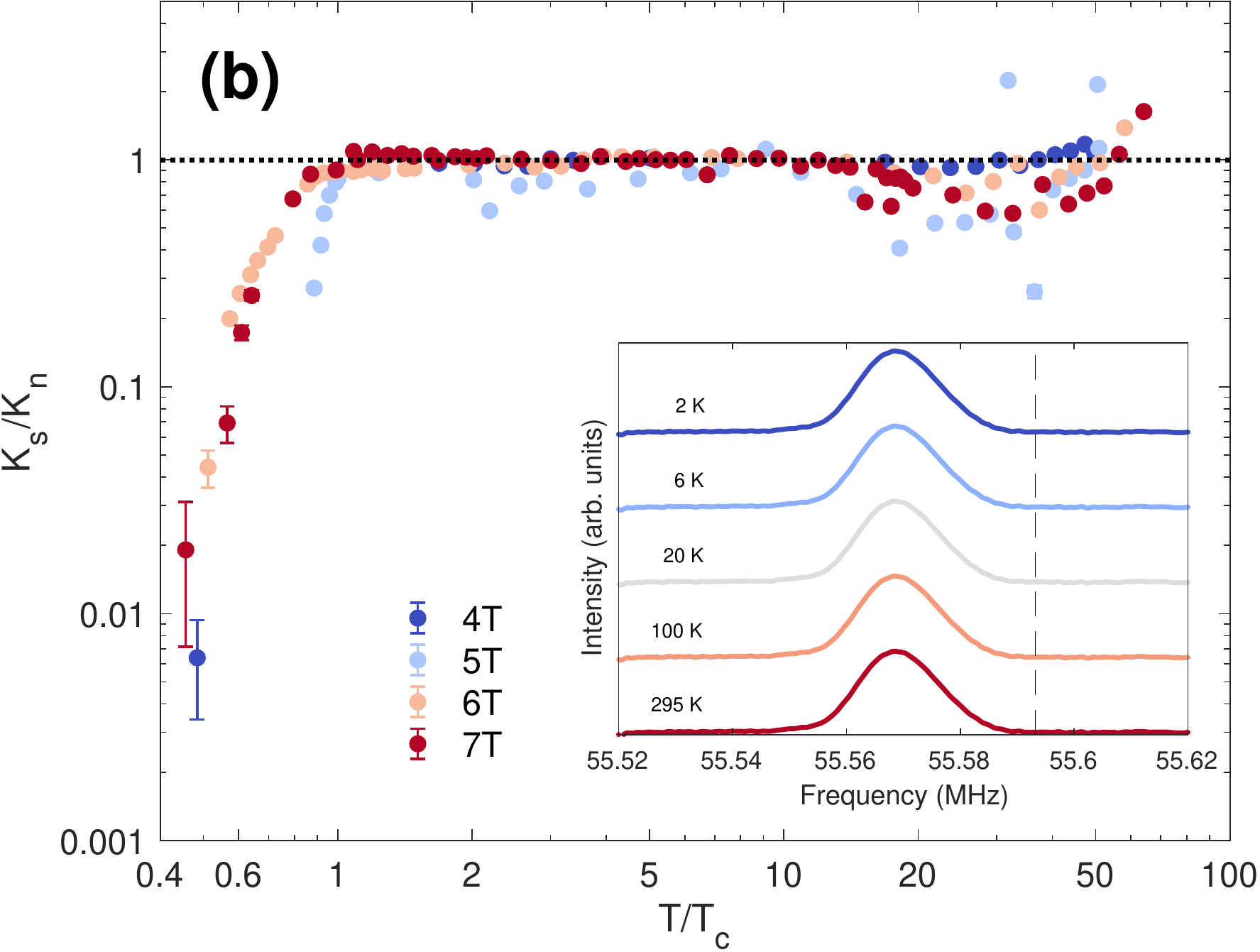}
   \caption{\label{fig:SC_evidence}Evidence of conventional BCS superconductivity
   in W$_{3}$Al$_{2}$C. (a) Below $T_{c}$, the relaxation rate ($T_{1}^{-1}$)
   can be described by a single-gap exponential function exp$(-\Delta/T)$.
   (b) The NMR shift $K_{s}$ (scaled by its normal-state value $K_{n}$ at
   $T_{c}$) remains constant down to $T_{c}$, but decreases sharply below
   it. Inset: $^{27}$Al NMR line shapes at selected temperatures measured
   in a 5-T applied magnetic field. The vertical line indicates the
   reference frequency $\nu_0$. 
   Above $T_c$, the lineshapes are practically temperature-independent.}
\end{figure*}

\section{\label{sec:results}Results and discussion}
\subsection{\label{ssec:NMR}\texorpdfstring{$^{27}$Al}{27Al} NMR in
the normal- and superconducting phases}

\begin{figure}[htbp]
\centering  
   \includegraphics[width=0.49\textwidth]{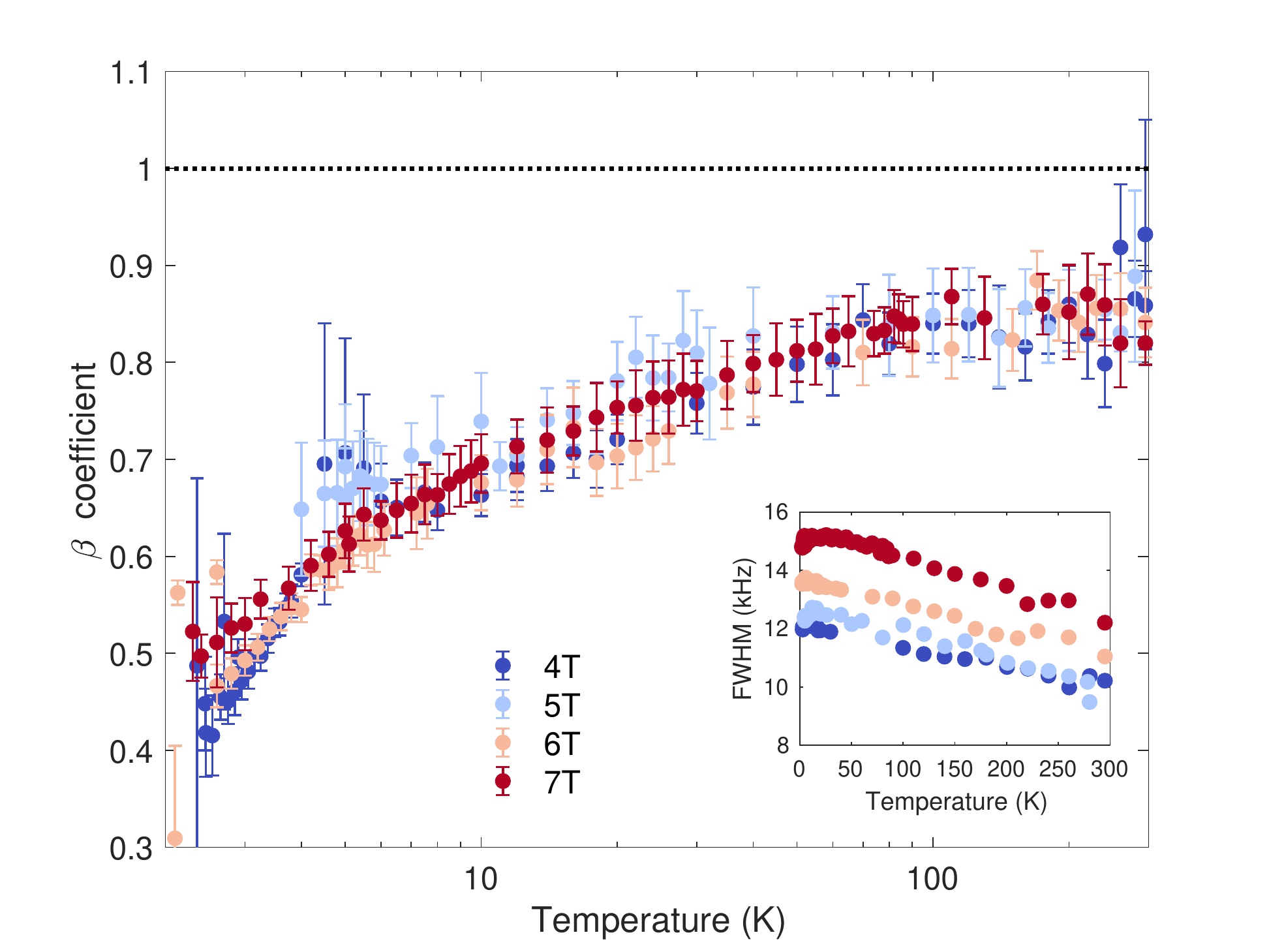}
   \caption{\label{fig:disorder}Temperature dependence of the stretching
   coefficient $\beta$. Inset: Temperature dependence of the linewidth.
   The behaviour of both parameters suggests the presence of intrinsic
   disorder in W$_{3}$Al$_{2}$C.}
\end{figure}

\begin{figure*}[htbp]
\centering
   \includegraphics[width=0.45\textwidth]{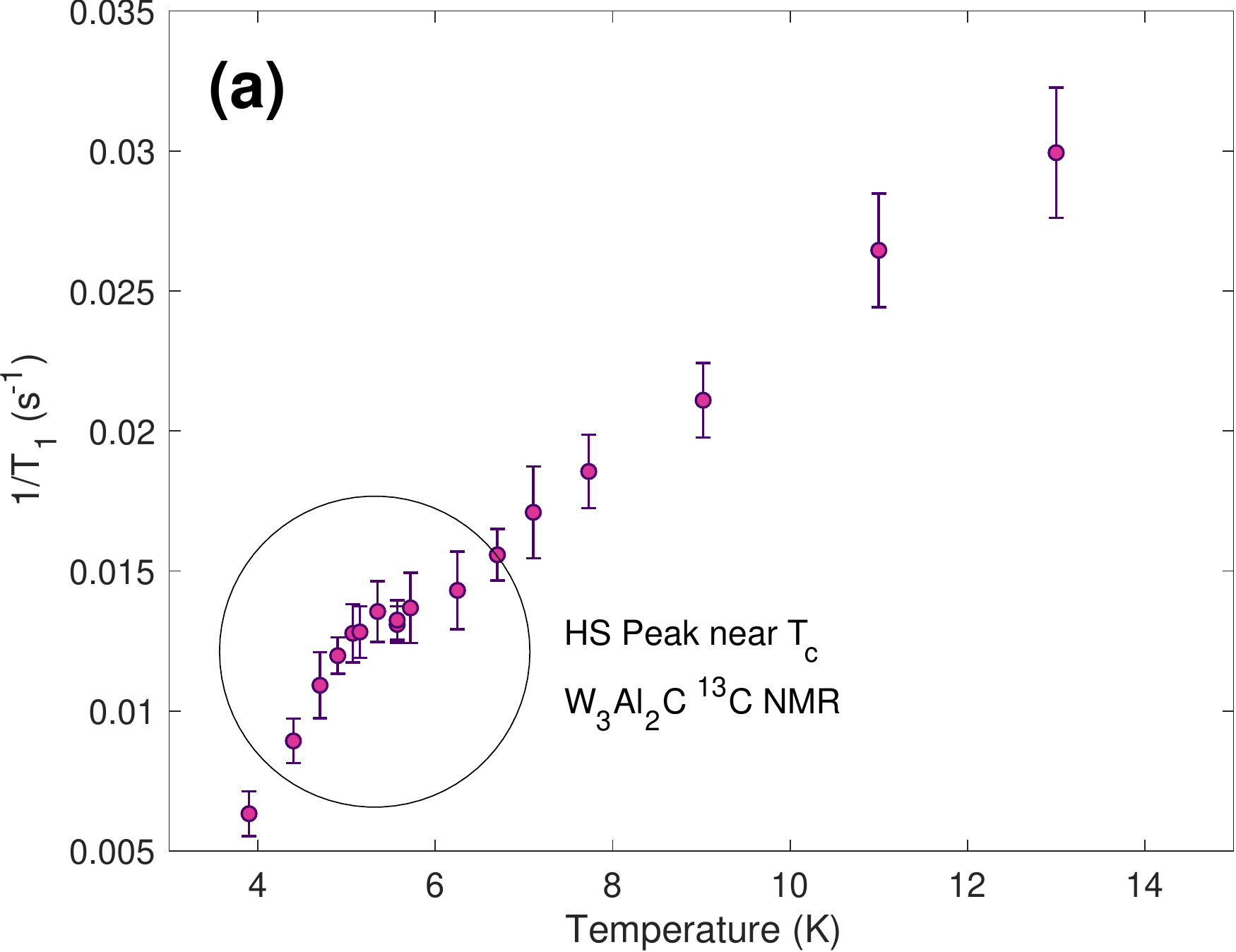}
   \hspace*{5mm}
   \includegraphics[width=0.45\textwidth,angle=0]{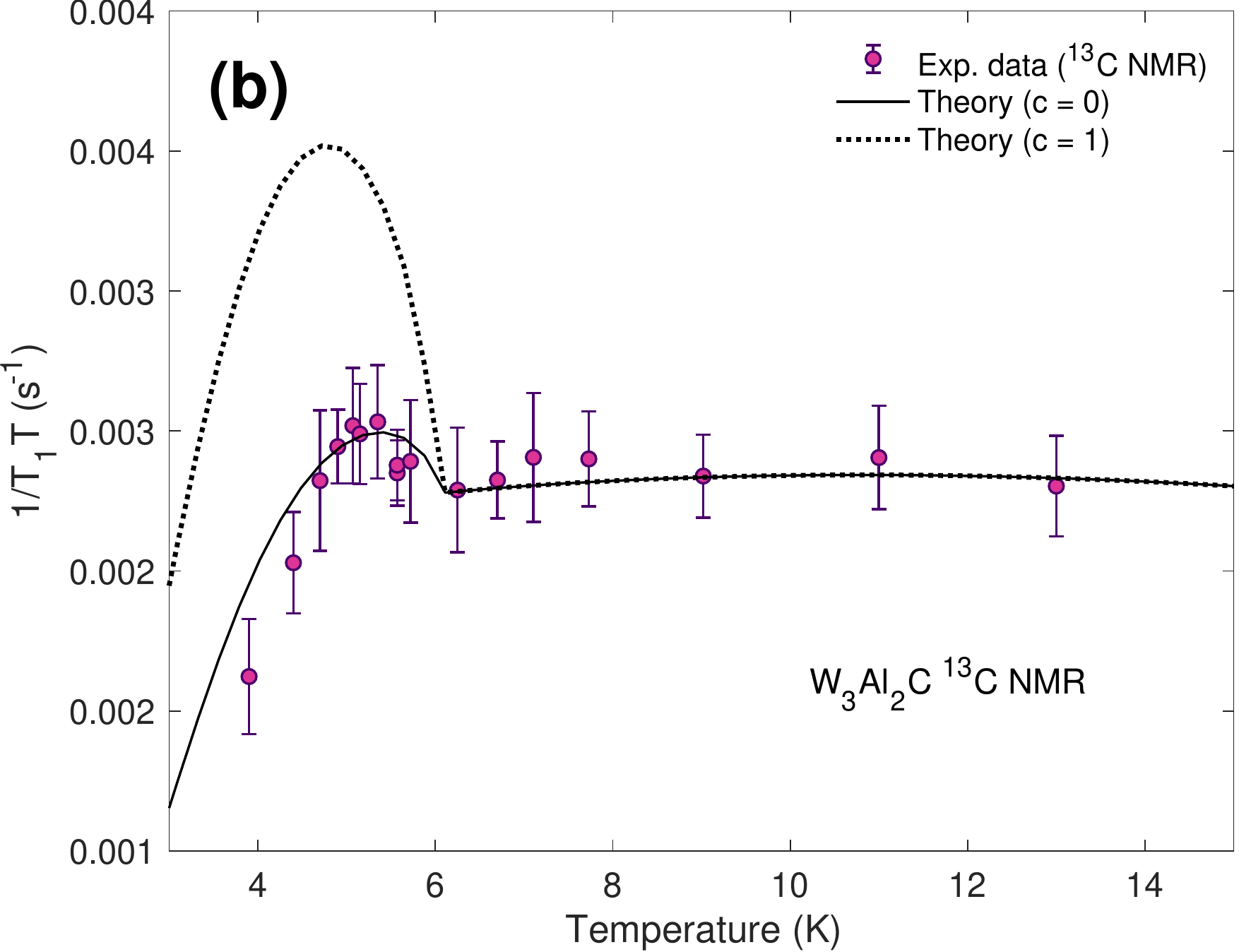}
   \caption{\label{fig:13C_NMR}$^{13}$C NMR relaxation data in W$_{3}$Al$_{2}$C.   
   Temperature dependence of the relaxation rate $T_{1}^{-1}$ (a) (the HS
   peak is circled in black) and of the Korringa product $(T_{1}T)^{-1}$ (b).
   Fits to equation~(\ref{eqn:HSpeak}) are also shown for the $c = 0$
   and $c = 1$ cases (see text for details).}
\end{figure*}

NMR is a powerful local technique for investigating the electronic
properties of materials, in particular, their electronic correlations,
magnetic order, and superconductivity \cite{Alloul2015}. Although the
magnetic field reduces the $T_c$ value, a high upper critical field
(above 50\,T~\cite{Gupta2021}), still allows us to explore the
superconducting behaviour of W$_{3}$Al$_2$C. In addition, we also
investigated \tco{the NMR response in the} 
normal state. 
The $^{27}$Al NMR
measurements were performed in different external magnetic fields, from
4 to 7\,T, corresponding to $T_c$ values between 5.9\,K and 4.6\,K. In
all cases, the $^{27}$Al NMR reference frequency $\nu_0$ was calculated
with respect to a standard Al(NO$_3$)$_3$ reference. The $T_1$ relaxation
times were calculated by using a stretched-exponential relaxation
model~\cite{McDowell1995,Narayanan1995}, whereby disorder is modelled
by a stretching coefficient $\beta$: 
%


\textcolor{black}{
\[\label{eq:T1_relax}
\frac{M_z(t)}{M_z^0} =  1-f\left(\frac{1}{35}e^{(-t/T_1)^{\beta}} + \frac{8}{45}e^{(-6t/T_1)^{\beta}} + \frac{50}{63}e^{(-6t/T_1)^{\beta}}
\right)\nonumber
\]
}
Here $M_z^0$ is the magnetization value at thermal equilibrium, while
$f$ reflects the efficiency of population inversion (ideally 2).
In general, the experimental NMR data --- shifts and relaxation rates
---  clearly reflect the two phases of W$_3$Al$_2$C: a normal metallic
phase above $T_c$ and a BCS-type superconducting phase below it.

\subsubsection{Metallic behaviour in the normal phase\hspace*{\fill}}
In the normal phase, we observe an almost ideal metallic behaviour for
all the applied magnetic fields (see figures~\ref{fig:T1T_all}(a) and
\ref{fig:SC_evidence}(b)). For instance, the relaxation-rate data in
figure~\ref{fig:T1T_all}(a) are well described by a power-law function
with a scaling exponent of 1.0(1), i.e., the relaxation rate is perfectly
proportional to temperature. We can also compute the Korringa product
$1/T_{1}T$ (here, only 0.017\,s$^{-1}$K$^{-1}$), which is proportional
to the electronic density of states at the Fermi level, $N(E_\mathrm{F})$ 
(see figure~\ref{fig:T1T_all}(b)).
As expected for an ideal metal, in the normal phase of W$_{3}$Al$_2$C,
$N(E_\mathrm{F})$ is practically constant with temperature and
independent of the applied \tco{magnetic} field, \tco{but} 
approximately thirty times
smaller than the Korringa product in 
metallic Al
(0.54 s$^{-1}$K$^{-1}$)~\cite{Carter1977,Carter1970}). These widely
differing relaxation rates are partly reflected also in 
the different Knight shifts: 450\,ppm in W$_{3}$Al$_2$C vs 1636\,ppm in
metallic Al~\cite{mehring1973high}, resulting in a $T_{1}TK^{2}$ of
$1.19\times10^{-5}$\,sK and $4.98\times10^{-6}$\,sK in  W$_{3}$Al$_2$C
and in metallic Al, respectively. The different $T_{1}TK^{2}$ values
suggest that, although W$_{3}$Al$_2$C is 
\tco{an ordinary metal in its}
normal phase, it could still exhibit weak electronic correlations.
The strong reduction in shift in W$_{3}$Al$_2$C suggests a very 
small contribution of $s$-electrons (responsible for the contact 
hyperfine interaction) at the Fermi energy. This is consistent with 
previous NMR studies of Al-based alloys \cite{Sandor2011} and with the 
electronic structure calculations of W$_{3}$Al$_2$C \cite{Ying2019}.
Indeed, the latter indicate that the $s$-electron band is shifted to 
higher energies, whereas the valence $d$-electron band of the W 
transition metal is prominent at the Fermi level.

\subsubsection{NMR evidence of BCS superconductivity in W$_3$Al$_2$C\hspace*{\fill}}
In the superconducting phase, the electronic properties of the system
can be described by the standard BCS theory, which predicts a gap ratio
$2\Delta_0/k_\mathrm{B}T_c = 3.53$ 
(see figure~\ref{fig:SC_evidence}(a)) and a reduction of the NMR shift
at low temperatures (see figure~\ref{fig:SC_evidence}(b)). Here, the
reduced shift reflects \tco{an} 
electron pairing 
\tco{forming} spin-compensated (typically $s$-wave) Cooper pairs. This result is consistent with a recent
muon-spin rotation ($\mu$SR) study of W$_{3}$Al$_2$C~\cite{Gupta2021}, 
where $s$-wave superconductivity with the same gap ratio was reported.
The superconducting gap value we find in W$_{3}$Al$_2$C is similar to the
\emph{average} SC gap reported in Mo$_{3}$Al$_2$C, where an NMR study
claimed a $2\Delta_0/k_\mathrm{B}T_c$ ratio of $2.8(2)$~\cite{Koyama2013},
while specific-heat measurements implied a ratio of 4.03~\cite{karki2010structure}.

\subsubsection{Detection of the Hebel-Slichter peak via $^{13}$C NMR\hspace*{\fill}}
%
%
According to the Korringa relation \cite{Korringa1950}, $T_1$ in metals
depends strongly on $N(E_\mathrm{F})$. Hence, upon entering the
superconducting state, we expect an exponential decay of $T_1$. Yet,
preceding this decay, just below $T_c$ the spin-lattice relaxation rate
first increases above its normal-state Korringa value, showing a so-called
coherence (or Hebel-Slichter --- HS) peak~\cite{Hebel1957,Hebel1959}.
Such increase in relaxation rate results partly from the enhanced
density of electronic states, which pile up near the SC gap
edges~\cite{hebel1959theory,Alloul2014}. Considering its key role in
the experimental confirmation of the BCS theory, the observation of a
Hebel-Slichter peak is recognized as a clear indication of BCS-type
$s$-wave superconductivity.

However, its absence does not necessarily rule out the possibility of a
standard $s$-wave pairing. This seems to be also the case for W$_3$Al$_2$C 
where, as illustrated in figure~\ref{fig:T1T_all}(a), we do not observe
the expected peak in the $^{27}$Al NMR relaxation rates below $T_c$.
Several hypotheses can be put forward to explain this experimental result.

One possibility, relevant to type-II superconductors, was suggested by
Goldberg and Weger~\cite{goldberg1968fluxoid}. Here, the total nuclear
spin-relaxation rate consists of the sum of two terms, the first of which
describes the relaxation in the normal-state vortex cores, while the
other captures the relaxation in the remaining superconducting volume.
This theory predicts that the contribution from the normal-state cores
is proportional to the external magnetic field $H$. To test it, we
measured the relaxation rates at different magnetic fields, in the
range from 4 to 7\,T. As can be seen in figure~\ref{fig:SC_evidence}(a),
the (scaled) relaxation times generally fall on the same exponential
curve, indicating that no field-dependent term arises from vortex cores.
Consequently, this hypothesis cannot explain the absence of an HS peak.

Another possibility is that the HS peak is suppressed by disorder,
caused by intrinsic defects and/or magnetic impurities. To investigate
the role of disorder, we studied the linewidths and the stretching
coefficient $\beta$. As shown in figure~\ref{fig:disorder}, $\beta$
decreases continuously as the temperature is lowered, while remaining
always smaller than 1. This indicates a distribution of relaxation rates,
reflecting the inequivalence of NMR sites, in turn due to the intrinsic
disorder (see, for instance, refs.~\cite{Shiroka2011,Shiroka2019}).
However, we find that both parameters depend only weakly on temperature
and show no discontinuities or an unusual behaviour near $T_{c}$. Thus,
we can rule out the possibility that disorder, although present, is
affecting the HS peak.

Yet another possibility is that, for quadrupolar nuclei such as the
spin-\nicefrac{5}{2} $^{27}$Al, also the quadrupole interaction
contributes to the spin-lattice relaxation, tending to smear out the
HS peak~\cite{li2016charge}. To test this possibility, we measured the
$^{13}$C NMR relaxation rate in W$_3$Al$_2$C. Since $^{13}$C is a
spin-\nicefrac{1}{2} nucleus, it represents a purely dipolar probe, not
undergoing any quadrupole interaction. As can be seen in
figure~\ref{fig:13C_NMR}(a), we indeed observe a small feature just
below $T_{c}$. Here, the exact $T_c$ value at 5\,T was determined by
means of a standard resonance-detuning experiment (data not shown). 

To model the feature near $T_{c}$, we begin with the usual expression 
for the HS peak~\cite{hebel1959theory}: 

\begin{equation}
\frac{R_{s}}{R_{n}}=\frac{2}{k_\mathrm{B}}
\int_{\Delta}^{\infty}\left[N_{s}^{2}(E)+M_{s}^{2}(E)\right]
f(E)[1-f(E)]\,\mathrm{d}E.
\label{eqn:HSpeak}
\end{equation}
Here, $R_{s}$ and $R_{n}$ are the relaxation rates in the
su\-per\-con\-duct\-ing and the normal state, respectively,
$M_{\mathrm{s}}(E) = N_{0} \Delta / \sqrt{E^{2}-\Delta^{2}}$ is the
anomalous density of states (DOS) due to the coherence
factor~\cite{hebel1959theory}, and
$N_{\mathrm{s}}(E) = N_{0} E / \sqrt{E^{2}-\Delta^{2}}$ is the DOS in
the superconducting state.

In our case, two modifications were made. Firstly, we chose to convolute 
$M_{s}(E)$ and $N_{s}(E)$ with a triangular broadening function $B(E)$,
characterized by a width $2\delta$ and a height $1/\delta$, with $\delta = 0.2\Delta$. 
Secondly, we model the degree of coherence by the parameter $c$, by
making the substitution $\Delta\rightarrow c\Delta$. \textcolor{black}{The 
conventional BCS theory \cite{Bardeen1957} \tcg{predicts} 
that, \tcg{depending on the details of the scattering operator,} the 
scattering matrix element can \tco{adopt} 
a positive or \tco{a} negative sign.
The \tcg{modelling} 
of the HS \tcg{peak} 
typically \tcg{requires} a positive sign, but in \tcg{other situations, 
e.g., in ultrasound absorption, 
the negative sign has to be used.}
\tcg{Here, by setting 
$c$ as} an adjustable parameter, one can model the case where the 
scattering matrix consists of both positive and negative components, 
which \tcg{may cause a (partly) suppression of} the HS peak.} After these two 
modifications, $N_{s}(E)$ and $M_{s}(E)$ can be written as \cite{Ding2016}:

\begin{eqnarray}
N_{s}(E) & =\int B\left(E^{\prime}-E\right) \frac{E^{\prime}}{\left(E^{\prime 2}-\Delta^{2}\right)^{1/2}}\,\mathrm{d} E^{\prime}, \\
M_{s}(E) & =\int B\left(E^{\prime}-E\right) \frac{c\Delta}{\left(E^{\prime 2}-\Delta^{2}\right)^{1/2}}\,\mathrm{d}E^{\prime}.
\end{eqnarray}

The fit results are shown in figure~\ref{fig:13C_NMR}(b). We find that
the fit obtained by fixing $c=1$ drastically overestimates the relaxation
rates below $T_{c}$. On the other hand, the fit obtained by fixing $c=0$
reproduces adequately the feature near $T_{c}$, although with a slightly
overestimated relaxation below $T_{c}$.

From the $^{13}$C NMR relaxation rates in the superconducting state and
the above analysis we \tco{infer with some confidence} 
that W$_{3}$Al$_{2}$C indeed exhibits
$s$-wave superconductivity, as evidenced by the 
\tco{observation} of an (albeit
reduced) HS peak. Yet, it appears that there are some decoherence effects,
both at and below $T_{c}$, most likely due to a strong electron-phonon
coupling in this material (see ref.~\cite{Ying2020} and the section
below). In general, the exact mechanism of such decoherence effects
is yet to be understood. 

\begin{figure}[!thb]
    \centering 
    \includegraphics[width=0.45\textwidth]{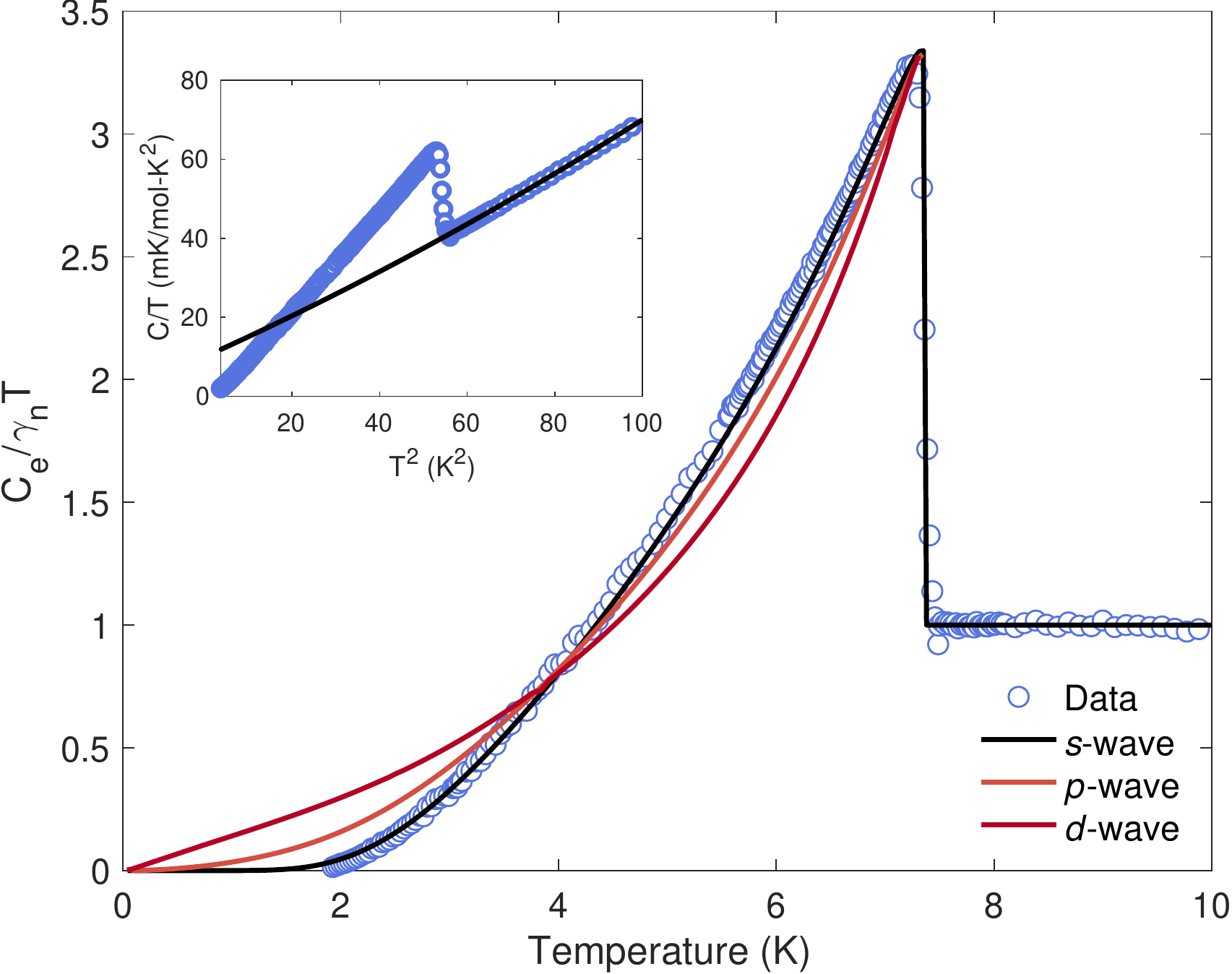}
    \caption{\label{fig:heatcapacity}Normalized electronic specific heat
    $C_\mathrm{e}/\gamma T$ for W$_3$Al$_2$C versus temperature. Inset:
    The measured specific heat $C/T$ as a function of $T^2$. The dashed
    line in the inset is a fit to $C/T = \gamma + \beta T^2 + \delta T^4 $,
    yielding $\gamma$ = 10(1)~mJ/mol-K$^2$, $\beta$ = 0.49(4)~mJ/mol-K$^4$,
    and $\delta$ = 1.0(3) $\times$ 10$^{-3}$~mJ/mol-K$^6$. The solid
    lines in the main panel are the electronic specific heat calculated
    by considering $s$-, $p$- and $d$-wave models. Specifc heat data were
    take from ref.~\cite{Ying2019}.} 
\end{figure}

\subsection{Electronic specific heat}
To get further insight into the superconducting state of W$_3$Al$_2$C,
\tco{the temperature dependence of} its electronic specific heat was evaluated and analyzed using different
models. After subtracting the phonon contribution from the measured
specific heat (see inset in figure~\ref{fig:heatcapacity}), the
electronic specific heat divided by the electronic specific-heat term,
i.e., $C_\mathrm{e}/\gamma T$, is obtained and presented in
figure~\ref{fig:heatcapacity}. The temperature-dependent
superconducting-phase contribution to the entropy was calculated
by means of the expression~\cite{Tinkham1996}:
\begin{equation}
	\label{eq:entropy}
	S(T) = -\frac{6\gamma_\mathrm{n}}{\pi^2 k_\mathrm{B}} \int^{\infty}_0 [f\mathrm{ln}f+(1-f)\mathrm{ln}(1-f)]\,\mathrm{d}\epsilon,
\end{equation}
\begin{table*}[]
\small
    \centering
    \setlength\tabcolsep{2mm}
    \begin{tabular}{p{36mm}p{28mm}p{21mm}p{21mm}p{21mm}p{21mm}}
    \toprule  
       Parameter & Rh$_{2}$Mo$_{3}$N & Mo$_{3}$Al$_{2}$C &  Li$_{2}$Pd$_{3}$B  & W$_{3}$Al$_{2}$C  & Li$_{2}$Pt$_{3}$B \\ \midrule
       $Z(M)$ & 42 & 42 &  46 & 74 & 78 \\ \midrule
       $T_{c}$\,(K) &  4.3 \cite{wei2016r}  &  9.0 \cite{bauer2010unconventional} & 7.2 \cite{takeya2007physical} & 7.6 \cite{Ying2019} & 2.6 \cite{takeya2007physical} \\
       $\gamma$\,(mJ/mol K$^{2}$) & 24.15 \cite{wei2016r} & 17.8 \cite{bauer2010unconventional} & 9.5  \cite{takeya2007physical}  &10 & 9.6 \cite{takeya2007physical} \\
       $\Delta C_{p}/\gamma T_{c}$ & 1.5 &  2.28 \cite{bauer2010unconventional} & 1.75 \cite{takeya2007physical}  &  2.3  & 0.75 \cite{takeya2007physical} \\
       $\xi$\,(nm) & 6.77 \cite{wei2016r} & 4.6 \cite{bauer2010unconventional}  & 7.6 \cite{takeya2007physical} & <5.7 \cite{Ying2019} & 11.8 \cite{takeya2007physical} \\
       $N(E_\mathrm{F})$ (states/eV/f.u.)& --- & 5.48 \cite{bauer2010unconventional}  &  2.24 \cite{Lee2005} & 2.37 \cite{Ying2019} & 2.9 \cite{Lee2005} \\
       $1/(T_{1}T)$(sK)$^{-1}$ &  --- & \textcolor{black}{ 0.34($^{27}$Al)\cite{Koyama2013} 0.075($^{27}$Al)\cite{bauer2010unconventional} 0.012($^{27}$Al)\cite{Kuo2012}}  & 0.045($^{11}$B)\cite{Nishiyama2007} & 0.017($^{27}$Al) & 0.034($^{11}$B)\cite{Nishiyama2007} \\ 
       \midrule
       $2\Delta/k_\mathrm{B}T_{c}$ ($C_{p}$) & 3.62 \cite{wei2016r} & 4.03 \cite{karki2010structure} & --- & 4.50  & non $s$-wave \\
       $2\Delta/k_\mathrm{B}T_{c}$ (NMR)& --- & 2.8 \cite{Koyama2013} & 2.2 \cite{nishiyama2005superconductivity} & 3.5 &  non $s$-wave \\
       $2\Delta/k_\mathrm{B}T_{c}$ ($\mu$SR) & 3.46 (dirty) \cite{TianReNb2018} & 2.59 \cite{bauer2014absence} & --- &  3.5 \cite{Gupta2021} & non $s$-wave \\
       & 3.84 (clean) \cite{TianReNb2018} & & & & \\
       \bottomrule
    \end{tabular}
    \caption{\label{tab:comparisontable}Comparison of the superconducting
    parameters of selected noncentrosymmetric superconductors with the
    chemical formula $M_{3}X_{2}Y$, where $M$ is a metal. In all the
    cited references,  \textcolor{black}{ $N(E_\mathrm{F})$ includes the contribution from SOC}.}
\end{table*}

where $\gamma_\mathrm{n}$ is the normal-state electronic specific-heat
coefficient, $f = (1+e^{E/k_\mathrm{B}T})^{-1}$ is the Fermi function and 
$E(\epsilon) = \sqrt{\epsilon^2 + \Delta^2(T)}$ is the excitation energy 
of quasiparticles, with $\epsilon$ the electron energies measured relative 
to the chemical potential (Fermi energy)~\cite{Tinkham1996,Padamsee1973}. 
Here, $\Delta(T)$ is the temperature dependent gap function, which in 
the BCS $s$-wave model can be written as 
$\Delta(T) = \Delta_0 \mathrm{tanh} \{ 1.82[1.018(T_\mathrm{c}/T-1)]^{0.51} \}$ 
\cite{Carrington2003}, with $\Delta_0$ the gap value at zero temperature. 
In case of $p$- and $d$-wave models, the temperature-dependent gap functions
are $\Delta(T)$sin$\theta$, and $\Delta(T)$cos2$\phi$, respectively, 
exhibiting point- and line nodes in the respective gap function. The
temperature-dependent electronic specific heat in the superconducting 
state can be calculated from $C_\mathrm{e} =T \frac{dS}{dT}$. 

The fit results using the above-mentioned models are shown by solid
lines in the main panel of figure~\ref{fig:heatcapacity}. We find that,
while the $s$-wave model fits the electronic specific heat data across
the entire temperature range, the $p$- and $d$-wave models deviate
significantly from the data, implying the absence of any gap nodes in
the superconducting state of W$_3$Al$_2$C. The fully-gapped $s$-wave model
gives a superconducting gap $\Delta_0 = 2.25(5)$\,$k_\mathrm{B}$$T_c$,
much larger than the expected weak-coupling BCS value (1.74\,$k_\mathrm{B}$$T_c$),
thus suggesting \emph{strongly-coupled}~\cite{Rainer1995} Cooper pairs
in W$_3$Al$_2$C. This is confirmed also by the specific-heat
discontinuity at $T_c$, i.e., $\Delta$$C$/$\gamma$$T_c$, here around
2.5--2.7, which is higher than 1.43 predicted by the BCS theory.

\subsection{Comparison with other members of the
\texorpdfstring{M$_{3}$X$_{2}$Y}{M3X2Y} family}
The superconducting properties of the $M_{3}X_{2}Y$ family of NCSCs are
summarized in table~\ref{tab:comparisontable}. The data are somewhat
challenging to interpret, because of the lack of a clear trend when
arranged in order of increasing atomic number $Z(M)$. Most likely this
reflects the fact that the superconducting behaviour of the $M_{3}X_{2}Y$ 
compounds is highly dependent on the details of their band structure,
which differs widely between members of the same family. Despite this
difficulty, we can still draw some conclusions.

First, we note that the SC gap values $2\Delta/k_\mathrm{B}T_c$ seem to
differ, depending on whether they were determined by means of NMR,
$\mu$SR, or $C_{p}$ measurements. In general, the gaps estimated via
NMR and $\mu$SR, which are both local-probe techniques, tend to be in
good agreement. Conversely, the SC gaps determined from $C_{p}$ data tend
to be higher than those calculated from NMR and $\mu$SR data, and
systematically indicate a strong electron-phonon coupling across the
entire family of $M_{3}X_{2}Y$ compounds.

\begin{figure}
    \centering
    \includegraphics[width=0.47\textwidth]{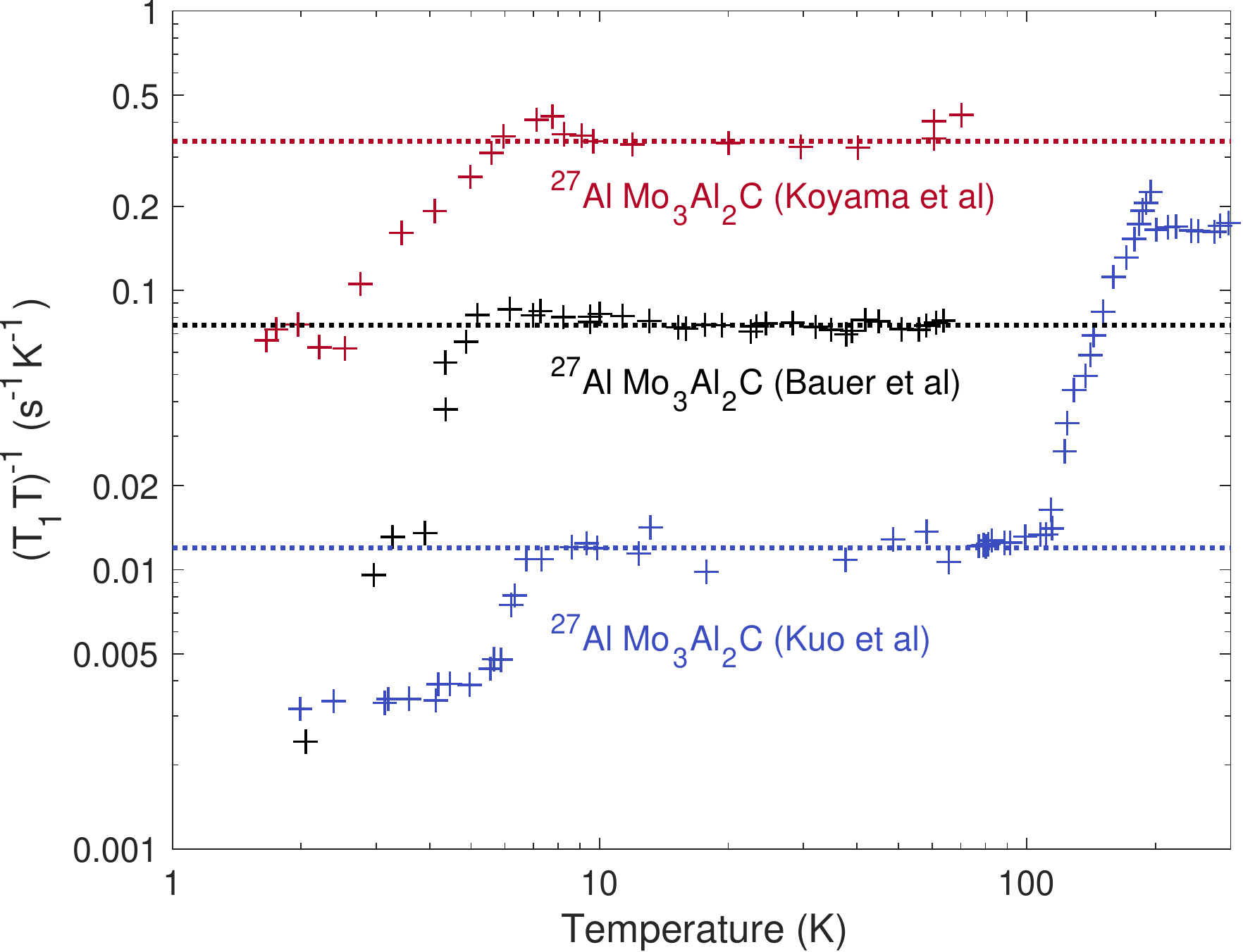}
    \caption{$1/T_1T$ vs temperature for Mo$_{3}$Al$_{2}$C as obtained from Koyama et al \cite{Koyama2013}, Bauer et al \cite{bauer2010unconventional} and Kuo et al \cite{Kuo2012}}
    \label{fig:T1T_Mo3Al2C}
\end{figure}

\textcolor{black}{
Secondly, we note that the $(T_{1}T)^{-1}$ values of W$_{3}$Al$_2$C are
consistent with \tcg{those of} 
the analog compound \tcg{Mo$_{3}$\-Al$_2$C} obtained by Bauer et 
al~\cite{bauer2010unconventional}. Since Mo$_{3}$Al$_{2}$C adopts the
same structure as W$_{3}$Al$_{2}$C, the only differences being that W is
less electronegative than Mo (1.7 vs 2.16 on the Pauling scale) and the
marginally larger atomic covalent radius of the W atoms compared to Mo
(150 vs 145\,pm), we expect the $N(E_\mathrm{F})$ of Mo$_{3}$Al$_{2}$C
and W$_{3}$Al$_{2}$C to be of a similar magnitude and, consequently,
their $(T_{1}T)^{-1}$ values to be comparable. \tcg{Considering that} 
the $N(E_\mathrm{F})$ of Mo$_{3}$Al$_{2}$C is twice as large 
\tcg{as that of} W$_{3}$Al$_{2}$C
(5.48 vs 2.37, see table~\ref{tab:comparisontable}), \tcg{this} 
results in an approximately fourfold increase in relaxation rate 
[$(T_{1}T)^{-1} = 0.075$ vs 0.017 (sK)$^{-1}$, see table~\ref{tab:comparisontable}]. 
On the other hand, the $(T_{1}T)^{-1}$ values obtained by Kuo et al \cite{Kuo2012} and Koyama et al \cite{Koyama2013} are wildly different (see figure \ref{fig:T1T_Mo3Al2C}). 
Recently, it was shown that the deliberate introduction of vacancies
(here Al) in $M_{3}X_{2}Y$ by soft chemical etching does not significantly influence the density of states (as from DFT calculations and heat-capacity
data), but instead \tcg{it} dramatically alters the strength of electron-phonon
coupling~\cite{Ying2020}. \tcg{Hence, the} presence of vacancies might 
\tcg{justify} the different $(T_{1}T)^{-1}$ values obtained by the 
different groups. Furthermore, since W$_{3}$Al$_{2}$C \tcg{intrisically} 
contains more Al-vacancies
than its Mo counterpart~\cite{Ying2020}, this might explain the BCS-type
character of the former.}

\textcolor{black}{Thirdly, our results in table~\ref{tab:comparisontable} 
support the observation \tcg{made} 
by Bauer et al~\cite{Bauer2012} that, in case of phonon-mediated superconductors, 
although a noncentrosymmetric crystalline structure is required \tcg{for 
the occurrence of unconventional superconductivity,} it \tco{does not} 
necessarily \tcg{imply it.} 
Another remarkable example of this is Mo$_{3}$P~\cite{Shang2019b} which,
as W$_{3}$Al$_{2}$C reported here, is a normal BCS-type superconductor, 
\tcg{despite being an NCSC.} 
\textcolor{black}{\tcg{Further,} even within the same family, isostructural 
compounds can display a different superconducting behavior. For instance, 
CePt$_3$Si and LaPt$_3$Si are isostructural ($P4mm$ space group) 
\tcb{and both display a large SOC}. \tcg{However, while} CePt$_3$Si exhibits unconventional 
superconductivity~\cite{bauer2004heavy,ribero2009magnetic}, LaPt$_3$Si 
does not~\cite{ribero2009magnetic}.}
Future studies could shed more light on this issue and clarify the link
between unconventional SC and the lack of a \tco{structural} inversion center.}

\section{Conclusion}
Extensive NMR- and specific-heat measurements in the non\-cen\-tro\-sym\-metric
W$_3$Al$_2$C superconductor, provide ample evidence about its weak
electron correlation, yet with a strong electron-phonon coupling.
Most importantly, we establish that W$_3$Al$_2$C shows a conventional 
BCS-type $s$-wave superconductivity. This, apparently unnoteworthy result, 
is surprising in view of the enhanced SOC of W atoms, whose 5$d$ orbitals 
dominate the density of states at the Fermi level. This is all the more
remarkable, if one considers that the analogous Mo$_3$Al$_2$C compound,
hosting the much lighter Mo atoms, is \tco{claimed to be} an 
unconventional superconductor. 
Such a counterintuitive outcome may be explained with the subtle role
played by spin-orbit coupling, as well as by its competition with the
electron bandwidth in the $M_{3}X_{2}Y$ family. Finally, by 
\tco{considering} the similar case of Mo$_3$P, we show that W$_3$Al$_2$C 
represents yet another example of a noncentrosymmetric material with conventional
normal-state- and superconducting properties, thus emphasizing 
the role of a noncentrosymmetric structure as a beneficial (but
not sufficient) condition in achieving unconventional superconductivity.

\section{Acknowledgements}
T.S.\ acknowledges support from the Natural
Science Foundation of Shanghai (Grants No.\ 21ZR1420500 and 21JC1402300).
This work was partially supported by the Schwei\-ze\-rische Na\-ti\-o\-nal\-fonds 
zur F\"{o}r\-de\-rung der Wis\-sen\-schaft\-lich\-en For\-schung, SNF
(Grant no.\ 200021-169455). 
Y.P.\ Qi was supported by the National Natural Science Foundation of
China (Grants No.\ U1932217 and 11974246) and the Science and Technology
Commission of Shanghai Municipality (19JC1413900).

\section{References}
\bibliography{W3Al2C_bib_v3}

\end{document}